\documentclass{cimento}

\usepackage{epsfig}
\title{Origin of the Ultrahigh-Energy Cosmic Rays and their  Spectral 
Break}
\author{
S.~Dado\from{ins:Technion}\ETC,
A.~Dar\from{ins:Technion}, 
A.~De R\'ujula\from{ins:CERN}}

\instlist{\inst{ins:Technion} Department of Physics,
Technion Israel Institute of Technology, Haifa 32000,
Israel, \inst{ins:CERN} Instituto de F\'isica Te\'orica (UAM/CSIC), Univ. 
Aut\'onoma de Madrid, Madrid and
CIEMAT, Madrid, Spain, CERN, 1211 Geneva 23, Switzerland,
Physics Department, Boston University, USA}
\PACSes{Cosmic Rays,~ 95.85.Ry, Gamma ray bursts,~98.70.Rz}

\begin{document}

\maketitle

\begin{abstract} 

The energy spectrum, composition and arrival directions of ultrahigh 
energy cosmic rays (UHECRs) with energy above the cosmic ray ankle, 
measured by the Pierre Auger Observatory, appear to be in conflict if 
their origin is assumed to be extragalactic. Their spectrum and 
composition, however, are those expected from Galactic UHECRs accelerated 
by highly relativistic jets such as those producing short hard gamma ray 
bursts (SHBs). If this alternative interpretation is correct, then the 
observed break in the energy spectrum of UHECRs around 50 EeV is the 
energy threshold for free escape of UHE iron nuclei from the Galaxy and 
not the Greisen-Zatsepin-Kuzmin (GZK) cutoff for protons, and the arrival 
directions of UHECR nuclei with energy above their UHE breaks must point 
back to their Galactic sources rather than to active galactic nuclei (AGN) 
within the GZK horizon.

\end{abstract}

Cosmic rays (CRs), discovered by Victor Hess \cite{ref:Hess1912} almost a 
century ago, have an observed spectrum extending from $E\!\leq\! 10^6$ eV 
to extremely high energies, $E\!\geq\! 10^{20}$ eV. At low energies the 
primary CRs contain all the stable elements. At very high energies, their 
all-particle spectrum has not been resolved into separate elements. Their 
energy spectrum is well represented by a broken power law $E^{-\beta}$, 
with $\beta\!\approx\!2.7$ above $\sim\!10$ GeV until the ``CR knee" at 
$\sim 3\times 10^{15}$ eV, where it steepens to $\beta\!\approx\!2.9$ up 
to a ``second knee" near $2\times 10^{17}$ eV where it changes to 
$\beta\!\approx\!3.3$. Above the "ankle" at $\sim\!3\times 10^{18}$ eV 
the ultrahigh energy cosmic ray (UHECR) flux has been accurately 
measured by the Fly's Eye High Resolution (HiRes) experiment 
\cite{ref:Abbasi2008, ref:Sokolsky2010} and the Pierre Auger Observatory 
(PAO) \cite{ref:Cronin2009, ref:Abraham2010a}. Its energy spectrum is well 
described by a power-law with $\beta\!\approx\! 2.7$ above until a 
"break" near $5\times 10^{19}$ eV where it changes to $\beta\!\approx\! 
4.3$, as shown in Fig.1, which shows the energy domain we are here 
concerned with.

While the origin of the CR knees of different elements is still debated, 
the CR ankle is generally identified as the energy beyond which the 
deflection of CRs in the Galactic magnetic field can neither isotropise 
them nor prolong significantly their residence time in the Galaxy, see 
e.g., \cite{ref:Dar1999, ref:Dar2008, ref:Hillas2005, ref:Wibig2005} and 
references therein, and \cite{ref:Berezinsky2006} for an alternative.

A free escape of UHECRs from the Galaxy implies that they essentially 
suffer an angular spread
$\langle\theta^2\rangle \ll\! 1$ by magnetic deflections on their way out
of the Galactic cosmic ray halo whose typical radius is $R_G\!\sim\!10$ 
kpc. For CRs of charge $Z$, this happens at an energy for which their 
Larmor radius, $R_L\!=\!E/Z\,e\, B_r$, becomes much larger than the 
coherence
length, $l_c $, of the random component of the Galactic magnetic field
$B_r \!\sim\! 3\, \mu$Gauss \cite{ref:Rand1989,ref:Beck2001} and their 
small
deflections $\delta\theta\simeq l_c/ R_L$ add up to:
\begin{equation}
\langle\theta\rangle \simeq \left [ {R_G\over l_c}\right]^{1/2}\,
\left[{l_c\over R_L}\right]\leq {\pi\over 2} \,.
\label{theta}
\end{equation}
For a typical $l_c\!\sim\! 0.1$ kpc, $R_L\!=\! E/e\, Z\, B$ and
$\theta\!=\!\pi/2$, Eq.~(\ref{theta}) yields a threshold energy for
escape, and consequently a spectral break at $E_{break}(A,Z)\!=\!Z\,
E_{break}(p)\!=\!1.8\,Z$ EeV. For Fe nuclei, $E_{break}(Fe)\!=\!46.8 $
EeV, which roughly coincides with the break-energy measured by 
HiRes \cite{ref:Abbasi2008, ref:Sokolsky2010} and PAO 
\cite{ref:Cronin2009, ref:Abraham2010a}.

The observed ultrahigh-energy (UHE) break at  $E\!\approx\! 5\times 
10^{19}$ eV 
was identified by both HiRes and PAO  as the so-called 
``GZK cutoff".  This  effective 
threshold for energy losses of CR protons by pion production  
in collisions with the cosmic microwave background (CMB) radiation, which
exponetially suppresses the extragalactic flux of UHECR protons with 
energy above $5\times 10^{19}$ eV, was  
predicted by Greisen \cite{ref:Greisen1966} and by     
Zatsepin and Kuzmin \cite{ref:Zatsepin1966} in 1966, right after the 
discovery of the CMB.

Further support for the identification of the UHE break 
with the GZK cutoff for UHECR protons came  from the arrival directions 
of UHECRs with energy above the GZK threshold observed in the early
PAO data \cite{ref:Abraham2007}: 
if the UHECRs are protons, half of those with $E\!\geq\! 
E_{GZK}$ must come  from distances $<\! 70$ Mpc. Indeed,
PAO reported that a large fraction of these
UHECRs (measured between 1 January 2004 and  31 August 2007)
had arrival directions pointing  back within $\leq\! 3.1\deg$ to 
active galactic nuclei (AGNs) closer than
$\sim\! 75$ Mpc, while the directions of those 
with smaller energies  were isotropic \cite{ref:Abraham2007,ref:Abraham2008}.

The conclusion that  most UHECRs with $E\!\geq\! E_{GZK}$
are protons was expected: extragalactic UHECR nuclei 
disintegrate in collisions with 
the infrared background radiations and the CMB, with a mean free 
path much shorter than that of UHE protons for $\pi$ production 
above the GZK threshold \cite{ref:Puget1976, ref:Epele1999,
ref:Dar2008, ref:Allard2008, ref:Aloisio2010}. 
 
However, this early evidence for a directional correlation with AGNs, 
obtained by PAO from a sample of 27 UHECRs was not 
present in a sample of an additional 42 events seen  
through 31 December 2009 and  has diminished significantly in the joint 
sample \cite{ref:Abreu2010}. 
In addition, the  HiRes collaboration reported   \cite{ref:Abbasi2010a}.
that their sample of 13 events with energy above 
57 EeV (1 EeV=$10^{18}$ eV), is
incompatible with directional correlation with AGNs
at 95\%.

Moreover, PAO recently reported the measured depth of shower maximum of 
UHECRs and its root-mean-square fluctuations, which indicate that the 
composition of UHECRs changes progressively with energy from 
proton-dominated below the CR ankle to Fe-dominated as one approaches the 
GZK cutoff \cite{ref:Cronin2009, ref:Abraham2010b}. The GZK cutoff for 
Fe-dominated 
composition is $\!\approx\! \rm A\!=\!56$ times larger than that for 
protons, 
$E_{GZK}(\rm Fe)\!\approx\! 3\times 10^{21}$ eV. Thus, the PAO composition 
of UHECRs seems to be in conflict with the identification of the 
UHE break at 50 EeV as the GZK cutoff of UHECR protons. Also 
the spectral shape around the UHE break seems not to be  compatible with 
that expected from the GZK cutoff \cite{ref:Aloisio2009}.
Note, however, that a proton dominated composition \cite{ref:Abbasi2010b}
and the spectrum of UHECRs  \cite{ref:Abbasi2010a}
that were measured by HiRes are those expected from 
extragalactic UHECRs \cite{ref:Dar2008}. 

All together, it appears that either 
the UHECRs are mainly extragalactic protons,  
the UHE break is due to the GZK cutoff
and the Fe-dominated composition of UHECRs 
near the GZK cutoff that was inferred by the PAO is not correct, 
or the UHECR composition becomes Fe-dominated near the
UHE break and the UHE break is not the GZK cutoff of UHECR protons.
This composition controversy, as well as the UHECR-AGN association 
controversy should be resolved experimentally. 
But, if the UHECR composition inferred by PAO  \cite{ref:Cronin2009, 
ref:Abraham2010b} turns out to be the correct one, is there a 
consistent  and simple explanation  for both the energy spectrum and 
composition measured by PAO?  

In this short paper we present such an explanation. 
We show that, with small modifications in the assumed
relative Galactic and extragalactic fluxes, the 
comprehensive theory of cosmic rays presented in \cite{ref:Dar2008}
correctly predicts
the energy spectrum and composition of the PAO UHECRs. 
All one has to do is to go back to the original
assumption that the UHECRs are dominantly of Galactic 
origin \cite{ref:Dar1999}.
We show that a rough knowledge of the properties of the Galactic
accelerators of UHECRs without an exact knowledge of their identity 
can reproduce the spectrum 
and composition of UHECRs which were reported by PAO. 

In \cite{ref:Dar2008} we posited that CRs are a mixture of Galactic 
and extragalactic fluxes, accelerated in gamma ray bursts 
(GRBs) \cite{ref:Dar1992}.
They are the GRB-ionized interstellar medium (ISM) collissionally accelerated by the highly 
relativistic jets of plasmoids (cannonballs) that produce 
Galactic and extragalactic GRBs, most of which
are  beamed away from Earth
\cite{ref:Dar1999}. The ones trapped in the Galactic magnetic field 
have produced its
CR halo. In a steady state, the escape rate from the CR halo 
equals its filling rate.
The two CR populations are steadily injected into the Galactic CR halo 
and the 
intergalactic medium (IGM) with roughly the same energy spectrum 
and composition. But they 
suffer different losses in the host galaxies of the GRBs and in 
the  IGM due to the different environments and residence times. 

The CR energy spectrum and composition
above the second knee reflect the A-dependent threshold 
energy (roughly proportional to A) for 
photo-dissociation of extragalactic CR nuclei in collisions with the 
CMB and the infrared background radiations during their 
long residence time in the IGM \cite{ref:Dar2008}. The second knee is 
the threshold for photo-dissociation of $^4$He. 
The CR composition changes progressively 
from that of low-energy CRs near the second
knee to almost a pure protons below the CR ankle,
as more heavy nuclei and their fragments disintegrate.    
The photo-disintegration of the primary nuclei and their fragments 
changes the power-law index of the all-particle energy spectrum from 
$\sim\!2.9$ below the second knee to $\sim\! 3.3$ 
above it. We do not discuss in detail this calculationally complex subject here  
(Dado and Dar in preparation) since we are focusing on the understanding
of UHECRs above the ankle. The Galactic 
component, whose residence time in the Galaxy is too short to 
imply a significant photo-disintegration in the ISM,
starts to dominate before the energy reaches the CR ankle.

Above  $E\!=\!E_{break}{}(p)\!\approx\! 
1.9$ EeV, UHECR protons are not isotropised and their 
free escape is not delayed by the Galactic magnetic 
field. Their flux predictably decreases with 
increasing energy beyond the proton UHE break. The CR nuclei of $^4$He, 
that at fixed particle energy are only 
slightly less abundant than protons  (by a factor $\sim\!0.8$), 
have a UHE break at $E_{break}(^4{\rm He})\!\approx\! 2\times 1.9$ EeV, beyond 
which Fe dominates the CR composition. The UHE Fe break is at 
$\approx\! 26\times 1.9\!\approx\! 50$ EeV. We shall interpret the UHE 
all-particle break as the UHE Fe 
break beyond which Fe CRs
are not isotropised nor confined. In order to validate this possibility,  
we proceed to derive the corresponding spectrum of UHECRs.

At the energies at which CR nuclei 
are isotropised by the Galactic magnetic field, their density 
is enhanced by their energy-dependent residence time in the Galaxy.
At relatively low (sub TeV) energies this time
 was empirically estimated \cite{ref:Swordy1990} to behave as
$\tau(E,Z)\!\propto\! (E/Z)^{-\beta_r}$ with $\beta_r\!\approx\! 
0.5\!\pm\!0.1$, 
yielding a CR number density \cite{ref:Dar2008}  
\begin{equation}
{dn_{_A}\over dE} \propto \tau(E,Z) {dn^{inj}_{_A}\over dE} \propto
     X(A,Z)\, A^{\beta-1}\, E^{-\beta}\,, 
\label{CRdensity}
\end{equation}
where $n^{inj}_{_A}$ are the injection rates of nuclei,  $X(A,Z)$ are their 
relative abundances in the ISM 
and  $\beta\!=\!\beta_{inj}+\beta_r$. 
For Fermi acceleration in highly relativistic jets, 
$\beta_{inj}\!=\!13/6$ for all CR nuclei \cite{ref:Dar2008}, while
$\beta_r$ is not known above the spectral knees.
If $\beta_r$ is $E$-independent, 
using its low-energy value one obtains \cite{ref:Dar2008}
a spectral index of Fe UHE nuclei   
$\beta\!=\!2.67\!\pm\! 0.1$, for
$E\!<\!E_{break}(\rm Fe)\!\sim \! 50$ EeV,
i.e.,
\begin{equation}
{dn_{_{Fe}}\over dE}\propto E^{-2.67\pm 0.1}\, .
\label{Ironspec}        
\end{equation}

Consider now the arrival of CR nuclei with
$E\! >\! E_{break}(A,Z)$ from Galactic sources. 
Their small deflections
by the  Galactic magnetic field 
along their path to Earth 
spread  their arrival directions according to  Eq.~(\ref{theta}) 
and their mean arrival times by  
\begin{equation}
\langle\tau_d (E,Z)\rangle \sim  {R_G \langle\theta^2\rangle\over 2\,c}
\label{tdelay}
\end{equation}
and their residence time in the Galaxy as a function of 
$E$ approaches rapidly their energy-independent free escape time,
$\langle\tau_r(E,Z)\rangle\!\sim \! (R_G/c)\, [1\! +\!\langle\theta^2\rangle/2]\rightarrow\! 
R_G/c$.
A distribution of $N_s$ Galactic 
transient sources of UHECRs that {\it isotropically emit} CRs
can produce a quasi-isotropic 
distribution of arrival directions provided their Galactic rate 
satisfies $\dot{N}_s\, \tau_d\! >\! 4\,/ \langle \theta^2\rangle$. 
Their spectral index, however, will remain $\beta\!=\!\beta_{inj}\!\approx\! 13/6$. 
    
In our theory \cite{ref:Dar2008} the injection of CRs is {\it narrowly 
beamed}:
CRs are accelerated by highly relativistic very narrow jets emitted in the birth 
or death of compact stars, in supernova explosions, in phase transitions
in compact stars, in their mergers, and in accretion 
episodes onto compact stars, all of which produce observable GRBs when their 
jets point towards Earth. 

For collimated sources Eq.~(\ref{theta}) implies that 
the probability for an UHECR to reach Earth is 
$\langle\theta^2\rangle/4 \!\propto\! E^{-2}$. Consequently, if 
the effective number of sources during $\tau_d(E)$ satisfies 
$N_{eff}\!=\!\dot N_s\, \tau_d(E) \!\ll\! 4/\langle\theta^2\rangle$,  the 
probability that the rays reach us during a time 
$\delta t\! \ll\! \tau_d(E)$ is $\propto\!E^{-2}$. The flux of 
UHECR nuclei beyond their $E_{break}(Z)\!=\!Z\, E_{break}(p)$ 
then satisfies
\begin{equation}
{dn_{_A}\over dE} \propto E^{-\beta_{inj}-2}\sim E^{-4.17}\,. 
\label{fbeyond} 
\end{equation}
This result is valid in the cannonball model of GRBs 
\cite{ref:Dar2004} where the jets 
have typical bulk-motion Lorentz factor $\gamma\!\sim\!10^3$ and the 
UHECRs are beamed into a cone with an opening angle 
$\theta\!\sim\!1/\gamma\!\sim\!10^{-3}$ much smaller than their 
angular spread by  Galactic 
magnetic deflections.  It is 
not valid in GRB fireball models with spherical ejecta or conical jets 
of opening angle much 
larger than the angular spread due to deflections by the Galactic 
magnetic field.

In Fig.1 we compare the PAO spectrum \cite{ref:Abraham2010a}
of UHECRs (multiplied by $E^3$ in Fig.2 for clarity)  and the approximate  power-law 
spectrum with the predicted indexes 
$\beta\!=\!3.3$ between the second knee and the 
$^4$He UHE break (Dado and Dar, in preparation),  
$\beta\!=\!2.67$ between this energy and the 
Fe break and $\beta\!=\!4.17$, as given 
in  Eqs.~(\ref{Ironspec},\ref{fbeyond}),
which follow from our current update of
\cite{ref:Dar2008}.  The results have been simplified by 
joining the power-law dependences at the ankle and UHE-break
transition points, and fitting the first of these transitions
to its observed energy.
The theoretical predictions agree well
with the PAO spectrum: a best fit yields  power-law
indixes 2.68 and 4.16  below and above the break, 
and 3.3 below the  ankle. 

A similar interpretation of the spectrum and composition of 
UHECRs has been proposed \cite{ref:Calvez2010}. 
It is based on the assumption that the origin of UHECRs 
is Galactic GRBs, as first suggested in 
\cite{ref:Dar1999}. Yet, we maintain that their derivation in 
\cite{ref:Calvez2010} of the spectrum  at energies 
above the UHE iron break is flawed \cite{ref:Comment2010}.

In conclusion, if the UHECRs with energy above the UHE break are mostly 
iron nuclei, as inferred from the PAO measurement, then the spectrum and 
composition of the UHECRs are those expected from CRs that are accelerated 
by the highly relativistic jets emitted in Galactic GRBs, most of which are 
mercifully beamed away from Earth. In particular, thehe UHE break in the 
spectrum of UHECRs 
around 50 EeV is not the GZK cutoff, but the energy threshold for 'free' 
escape of UHE Fe nuclei from the Galaxy. The energy spectrum of UHECRs 
above the UHE break is a trivial consequence of the energy 
dependence of the magnetic deflection of Galactic UHE Fe nuclei: It is the 
steepening by two units of the spectrum at the break, Eq.(\ref{fbeyond}) 
that reflects the ``rigidity" of a UHECR trajectory in the randomly 
directed domains of the Galactic magnetic field. Finally
the UHECR nuclei above their respective UHE breaks should point back 
towards young remnants of Galactic GRBs. These may be supernova 
remnants, magnetars, young neutron stars and accreting compact objects 
in close binaries (the expected angular-clustering properties of 
UHECRs will be discussed elsewhere).

If the UHECRs are extragalactic protons, as implied by the Fly's Eye HiRes 
observations, then the UHE break near 50 EeV is the GZK cutoff, and the 
UHECRs must be accompanied by the UHE neutrinos and photons from the decay 
of the charged and neutral GZK pions. Their expected spectral index between 
the CR ankle and the GZK cutoff is their injection index below the CR 
ankle, i.e. $\beta\!=\!3.2\!-\!\beta_r\!\approx\!2.7\!\pm\!0.1$, their 
spectrum above the UHE break is the GZK spectrum, and their arrival 
directions should point towards their nearby extragalactic sources.

\begin{figure}[]
\centering
\epsfig{file=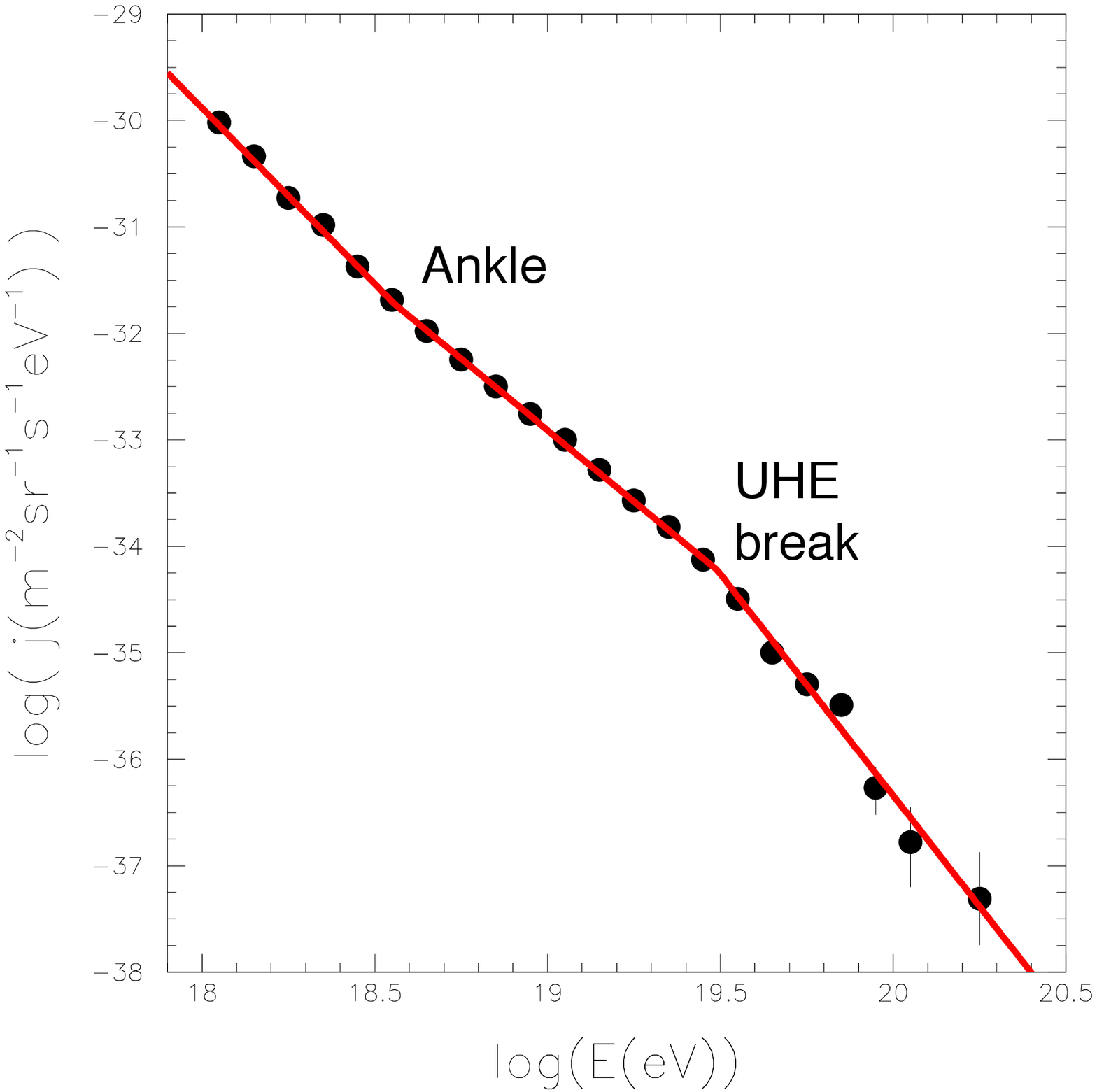,width=14cm,height=14cm} 
\caption{Comparison between the predicted slopes of the broken power-law 
spectrum of UHECRs and the PAO data \cite{ref:Abraham2010a}. 
The overall normalization and the energy of the cosmic ray ankle 
which depend on poorly known astrophysical
parameters, were adjusted by a best fit 
to the data.} 
\end{figure}

\begin{figure}[]
\centering
\epsfig{file=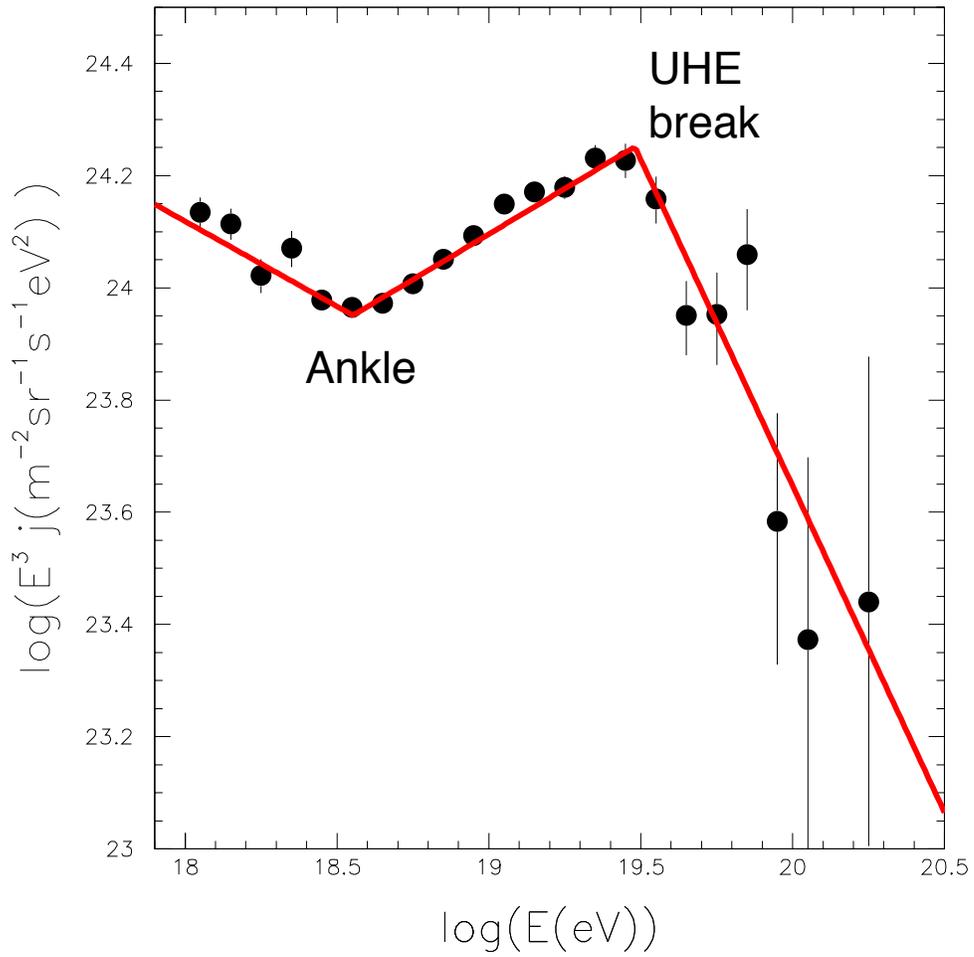,width=14cm,height=14cm}
\caption{The predicted broken power-law 
spectrum of UHECRs, compared to the  
PAO data \cite{ref:Abraham2010a}, both
multiplied by $E^3$.}
\end{figure}

\end{document}